\documentstyle[12pt,epsfig]{ioplppt}          % use this for preprint style

\begin{document} 
\jl{1}
\date{August 15th, 1996}

\title{The dynamics of a Genetic Algorithm for a simple learning
problem}[The dynamics of a GA for a simple learning problem]

\author{Magnus Rattray\ftnote{3}{email: rattraym@cs.man.ac.uk} and
  Jonathan L Shapiro\ftnote{1}{email: jls@cs.man.ac.uk}}

\address{Computer Science Department, University of Manchester, Oxford
Road, Manchester M13 9PL, UK}

\address{To appear in Journal of Physics A. Accepted: August 15, 1996}

\begin{abstract} 

A formalism for describing the dynamics of Genetic Algorithms (GAs)
 using methods from statistical mechanics is applied to the problem of
 generalization in a perceptron with binary weights. The dynamics are
 solved for the case where a new batch of training patterns is
 presented to each population member each generation, which
 considerably simplifies the calculation. The theory is shown to agree
 closely to simulations of a real GA averaged over many runs,
 accurately predicting the mean best solution found. For weak
 selection and large problem size the difference equations describing
 the dynamics can be expressed analytically and we find that the
 effects of noise due to the finite size of each training batch can be
 removed by increasing the population size appropriately.  If this
 population resizing is used, one can deduce the most computationally
 efficient size of training batch each generation. For independent
 patterns this choice also gives the minimum total number of training
 patterns used. Although using independent patterns is a very
 inefficient use of training patterns in general, this work may also
 prove useful for determining the optimum batch size in the case where
 patterns are recycled.

\end{abstract}

%
%  Uncomment out if preprint format required
%
%\pacs{00.00, 20.00, 42.10}
%\maketitle            

\section{Introduction}

Genetic Algorithms (GAs) are adaptive search techniques, which can be
used to find low energy states in poorly characterized,
high-dimensional energy landscapes~\cite{Gold,Holl}. They have already
been successfully applied in a large range of domains~\cite{Handbook}
and a review of the literature shows that they are becoming
increasingly popular. In particular, GAs have been used in a number of
machine learning applications, including the design and training of
artificial neural networks~\cite{Fitz,Sch,Yao}.

In the simple GA considered here, each population member is
represented by a genotype, in this case a binary string, and an
objective function assigns an energy to each such genotype. A
population of solutions evolves for a number of discrete generations
under the action of genetic operators, in order to find low energy
(high fitness) states. The most important operators are selection,
where the population is improved through some form of preferential
sampling, and crossover (or recombination), where population members
are mixed, leading to non-local moves in the search space. Mutation is
usually also included, allowing incremental changes to population
members. GAs differ from other stochastic optimisation techniques,
such as simulated annealing, because a population of solutions is
processed in parallel and it is hoped that this may lead to
improvement through the recombination of mutually useful features from
different population members.

A formalism has been developed by Pr\"{u}gel-Bennett, Shapiro and
Rattray which describes the dynamics of a simple GA using methods from
statistical mechanics~\cite{adam,PBS,PBS2,Ratt}. This formalism has
been successfully applied to a number of simple Ising systems and has
been used to determine optimal settings for some of the GA search
parameters~\cite{PBS3}. It describes problems of realistic size and
includes finite population effects, which have been shown to be
crucial to understanding how the GA searches. The approach can be
applied to a range of problems including ones with multiple optima,
and it has been shown to predict simulation results with high
accuracy, although small errors can sometimes be detected.

Under the statistical mechanics formalism, the population is described
by a small number of macroscopic quantities which are statistical
measures of the population.  Statistical mechanics techniques are used
to derive deterministic difference equations which describe the
average effect of each operator on these macroscopics.  Since the
dynamics of a GA is to be modelled by the average dynamics of an
ensemble of GAs, it is important that the quantities which are used to
describe the system are robust and self-averaging.  The macroscopics
which have been used are the cumulants of some appropriate quantity,
such as the energy or the magnetization, and the mean correlation
within the population, since these are robust statistics which average
well over different realizations of the dynamics. There may be small
systematic errors, since the difference equations for evolving these
macroscopics sometimes involve nonlinear terms which may not
self-average, but these corrections are generally small and will be
neglected here.

The statistical mechanics theory is distinguished by the facts that a
macroscopic description of the GA is used and that the averaging is
done such that fluctuations can be included in a systematic way.  Many
other theoretical approaches are based on the intuitive idea that
above average fitness building blocks are preferentially sampled by
the GA, which, if they can be usefully recombined, results in highly
fit individuals being produced~\cite{Gold,Holl}. Although this may be
a useful guide to the suitability of particular problems to a GA, it
is difficult to make progress towards a quantitative description for
realistic problems, as it is difficult to determine which are the
relevant building blocks and which building blocks are actually
present in a finite population. This approach has led to false
predictions of problem difficulty, especially when the dynamic nature
of the search is ignored~\cite{Forr,Greff}.  A rigorous approach
introduced by Vose \etal describes the population dynamics as a
dynamical system in a high-dimensional Euclidean space, with each
genetic operator incorporated as a transition
tensor~\cite{Vose1,Vose2}. This method uses a microscopic description
and is difficult to apply to specific problems of realistic size due
to high-dimensionality of the equations of motion.  More recently, a
number of results have been derived for the performance of a GA on a
class of simple additive problems \cite{Baum,Muhl,Theirens}. These
approaches use a macroscopic description, but assume a particular form
for the distribution of macroscopics which is only applicable in large
populations and for a specific class of problem. It is difficult to
see how to transfer the results to other problems where finite
population effects cannot be ignored.

Other researchers have introduced theories based on averages.  A
description of GA dynamics in terms of the evolution of the parent
distribution from which finite populations are sampled was produced by
Vose and Wright~\cite{Vose95}. This microscopic approach provides a
description of the finite population effects which is elegant and
correct. However, like other microscopic descriptions it is difficult
to apply to specific realistic problems due to the enormous
dimensionality of the system. Macroscopic descriptions can result in
low-dimensional equations which can be more easily studied.  Another
formalism based on the evolution of parent distributions was developed
by Peck and Dhawan~\cite{Peck}, but they did not use the formalism to
develop equations describing finite population dynamics.

The importance of choosing appropriate quantities to average is
well-known in statistical physics, but does not seem to be widely
appreciated in genetic algorithm theory. In particular, many authors
use results based on properties of the {\em average} probability
distribution; this is insensitive to finite-population fluctuations
and only gives accurate results in the infinite population
limit. Thus, many results are only accurate in the infinite population
limit, even though this limit is not taken explicitly. For example,
Srinivas and Patnaik~\cite{Sri} and Peck and Dhawan~\cite{Peck} both
produce equations for the moments of the fitness distribution in terms
of the moments of the initial distribution. These are moments of the
average distribution. Consequently, the equations do not correctly
describe a finite population and results presented in these papers
reflect that. Other attempts to describe GAs in terms of population
moments (or schema moments or average Walsh coefficients) suffer from
this problem.  Macroscopic descriptions of population dynamics are
also widely used in quantitative genetics (see, for example,
reference~\cite{Falconer}).  In this field the importance of
finite-population fluctuations is more widely appreciated; the
infinite population limit is usually taken explicitly.  Using the
statistical mechanics approach, equations for fitness moments which
include finite-population fluctuations can be derived by averaging the
cumulants, which are more robust statistics.

Here, the statistical mechanics formalism is applied to a simple
problem from learning theory, generalization of a rule by a perceptron
with binary weights. The perceptron learns from a set of training
patterns produced by a teacher perceptron, also with binary weights. A
new batch of training patterns are presented to each population member
each generation which simplifies the analysis considerably, since
there are no over-training effects and each training pattern can be
considered as statistically independent. Baum \etal have shown that
this problem is similar to a paramagnet whose energy is corrupted by
noise and they suggest that the GA may perform well in this case,
since it is relatively robust towards noise when compared to local
search methods~\cite{Baum}. The noise in the training energy is due to
the finite size of the training set and is a feature of many machine
learning problems~\cite{Fitz}.

We show that the noise in the training energy is well approximated by
a Gaussian distribution for large problem size, whose mean and
variance can be exactly determined and are simple functions of the
overlap between pupil and teacher. This allows the dynamics to be
solved, extending the statistical mechanics formalism to this simple,
yet non-trivial, problem from learning theory. The theory is compared
to simulations of a real GA averaged over many runs and is shown to
agree well, accurately predicting the evolution of the cumulants of
the overlap distribution within the population, as well as the mean
correlation and mean best population member. In the limit of weak
selection and large problem size the population size can be increased
to remove finite training set effects and this leads to an expression
for the optimal training batch size.

\section{Generalization in a perceptron with binary weights}

A perceptron with Ising weights $w_i \in \{-1,1\}$ maps an Ising
training pattern $\{\zeta_i^\mu\}$ onto a binary output,
\begin{equation}
        O^\mu = {\rm Sgn}\left(\sum_{i=1}^N w_i \zeta_i^\mu\right)
                \qquad {\rm Sgn}(x)=\cases{ 1&for $x \geq 0$ \\ -1&for
                $x<0$\\}
\end{equation}
where $N$ is the number of weights. Let $t_i$ be the weights of the
teacher perceptron and $w_i$ be the weights of the pupil. The
stability of a pattern is a measure of how well it is stored by the
perceptron and the stabilities of pattern $\mu$ for the teacher and
pupil are $\Lambda_t^\mu$ and $\Lambda_w^\mu$ respectively,
\begin{equation}
        \Lambda_t^\mu = \frac{1}{\sqrt{N}}\sum_{i=1}^N t_i \zeta_i^\mu
        \hspace{1cm} \Lambda_w^\mu = \frac{1}{\sqrt{N}}\sum_{i=1}^N
        w_i \zeta_i^\mu
\end{equation}
The training energy will be defined as the number of patterns the
pupil misclassifies,
\begin{equation}
        E = \sum_{\mu=1}^{\lambda N}
                \Theta(-\Lambda_t^\mu\Lambda_w^\mu) \qquad
                \Theta(x)=\cases{ 1&for $x \geq 0$ \\ 0&for $x<0$\\}
\label{def_E}
\end{equation}
where $\lambda N$ is the number of training patterns presented and
$\Theta(x)$ is the Heaviside function. In this work a new batch of
training examples is presented each time the training energy is
calculated.

For large $N$ it is possible to calculate the entropy of solutions
compatible with the total training set and there is a first-order
transition to perfect generalization as the size of training set is
increased~\cite{Gyorg,Somp}. This transition occurs for $O(N)$
patterns and beyond the transition the weights of the teacher are the
only weights compatible with the training set. In this case there is
no problem with over-training to that particular set, although a
search algorithm might still fail to find these weights. The GA
considered here will typically require more than $O(N)$ patterns,
since it requires an independent batch for each energy evaluation, so
avoiding any possibility of over-training.

Define $R$ to be the overlap between pupil and teacher,
\begin{equation}
        R = \frac{1}{N}\sum_{i=1}^N w_i t_i \label{def_R}
\end{equation}
We choose $t_i = 1$ at every site without loss of generality.  If a
statistically independent pattern is presented to a perceptron, then
for large $N$ the stabilities of the teacher and pupil are Gaussian
variables each with zero mean and unit variance, and with covariance
$R$,
\begin{equation}
        p(\Lambda_t,\Lambda_w) =
        \frac{1}{2\pi\sqrt{1-R^2}}\exp\!\left(\frac{-(\Lambda_t^2 -
        2R\Lambda_t\Lambda_w + \Lambda_w^2)}{2(1-R^2)}\right)
        \label{stab_dist}
\end{equation}
The conditional probability distribution for the training energy given
the overlap is,
\begin{equation}
        p(E|R) = \left\langle \delta\!\left(E - \sum_{\mu =
        1}^{\lambda N} \Theta(-\Lambda_t^\mu\Lambda_w^\mu)
        \right)\right\rangle_{\!\!\{\Lambda_t^\mu,\Lambda_w^\mu\}}
        \label{def_p}
\end{equation}
where the brackets denote an average over stabilities distributed
according to the joint distribution in equation~(\ref{stab_dist}). The
logarithm of the Fourier transform generates the cumulants of the
distribution and using the Fourier representation for the delta
function in $p(E|R)$ one finds,
\begin{eqnarray}
        \hat{\rho}(-\i t|R) \:&= \:\int_{-\infty}^\infty \!\!\!\d E\,
                        p(E|R) \, \e^{tE} \nonumber \\ &=
                        \:\left\langle \prod_{\mu=1}^{\lambda N}
                        \exp\left[t\Theta(-\Lambda_t^\mu\Lambda_w^\mu)\right]
                        \right\rangle \nonumber \\ &= \:\left(1 +
                        \frac{1}{\pi}(\e^t-1)\cos^{-1}(R)
                        \right)^{\!\lambda N}
\end{eqnarray}
The logarithm of this quantity can be expanded in $t$, with the
cumulants of the distribution given by the coefficients of the
expansion. The higher cumulants are $O(\lambda N)$ and it turns out
that the shape of the distribution is not critical as long as
$\lambda$ is $O(1)$. A Gaussian distribution will be a good
approximation in this case,
\begin{equation}
        p(E|R) = \frac{1}{\sqrt{2\pi\sigma^2}}\exp\!\left(\frac{-(E -
        E_{\rm g}(R))^2}{2\sigma^2}\right)
\label{def_p_g}
\end{equation}
where the mean and variance are,
\begin{eqnarray} 
        E_{\rm g}(R) \; = \; \frac{\lambda N}{\pi} \cos^{-1}(R)
        \label{<E|R>} \\ \sigma^2 \; = \; \frac{\lambda
        N}{\pi}\cos^{-1}(R)\left(1 -
        \frac{1}{\pi}\cos^{-1}(R)\right)\label{sigma}
\end{eqnarray}
Here, $E_{\rm g}(R)$ is the generalization error, which is the
probability of misclassifying a randomly chosen training example. The
variance expresses the fact that there is noise in the energy
evaluation due to the finite size of the training batch.

\section{Modelling the Genetic Algorithm}
\subsection{The Genetic Algorithm}

Initially, a random population of solutions is created, in this case
Ising weights of the form $\{w_1,w_2\ldots,w_N\}$ where the alleles
$w_i$ are the weights of a perceptron. The size of the population is
$P$ and will usually remain fixed, although a dynamical resizing of
the population is discussed in section~\ref{sec_rescale}. Under
selection, new population members are chosen from the present
population with replacement, with a probability proportional to their
Boltzmann weight. The selection strength $\beta$ is analogous to the
inverse temperature and determines the intensity of selection, with
larger $\beta$ leading to a higher variance of selection
probabilities~\cite{Maza,PBS}.  Under standard uniform crossover, the
population is divided into pairs at random and the new population is
produced by swapping weights at each site within a pair with some
fixed probability. Here, bit-simulated crossover is used, with new
population members created by selecting weights at each site from any
population member in the original population with equal
probability~\cite{Sysw}.  In practice, the alleles at every site are
completely shuffled within the population and this brings the
population straight to the fixed point of standard crossover. This
special form of crossover is only practicable here because crossover
does not change the mean overlap between pupil and teacher within the
population. Standard mutation is used, with random bits flipped
throughout the population with probability $p_{\rm m}$.
    
Each population member receives an independent batch of $\lambda N$
examples from the teacher perceptron each generation, so that the
relationship between the energy and the overlap between pupil and
teacher is described by the conditional probability defined in
equation~(\ref{def_p}). In total, $\lambda N\!\times\!PG$ training
patterns are used, where $G$ is the total number of generations and
$P$ is the population size (or the mean population size).

\subsection{The Statistical Mechanics formalism}  

The population will be described in terms of a number of macroscopic
variables, the cumulants of the overlap distribution within the
population and the mean correlation within the population.  In the
following sections, difference equations will be derived for the
average change of a small set of these macroscopics, due to each
operator. A more exact approach considers fluctuations from mean
behaviour by modelling the evolution of an ensemble of populations
described by a set of order parameters~\cite{adam}. Here, it is
assumed that the dynamics average sufficiently well so that we can
describe the dynamics in terms of deterministic equations for the
average behaviour of each macroscopic. This assumption is justified by
the excellent agreement between the theory and simulations of a real
GA, some of which are presented in section~\ref{sec_sim}. Once
difference equations are derived for each macroscopic, they can be
iterated in sequence in order to simulate the full dynamics.

Notice that although we follow information about the overlap between
teacher and pupil, this is of course not known in general. The only
feedback available when training the GA is the training energy defined
in equation~\ref{def_E}. Selection acts on this energy, and it is
therefore necessary to average over the noise in selection which is
due both to the stochastic nature of the training energy evaluation
and of the selection procedure itself.

Finite population effects prove to be of fundamental importance when
modelling the GA. A striking example of this is in selection, where an
infinite population assumption leads to the conclusion that the
selection strength can be set arbitrarily high in order to move the
population to the desired solution. This is clearly nonsense, as
selection could never move the population beyond the best existing
population member. Two improvements are required to model selection
accurately; the population should be finite and the distribution from
which it is drawn should be modelled in terms of more than two
cumulants, going beyond a Gaussian approximation~\cite{PBS}. The
higher cumulants play a particularly important role in selection which
will be described in section~\ref{sec_selapp}~\cite{PBS2}.

The higher cumulants of the population after bit-simulated crossover
are determined by assuming the population is at maximum entropy with
constraints on the mean overlap and correlation within the population
(see \ref{max_ent}). The effect of mutation on the mean overlap and
correlation only requires the knowledge of these two macroscopics, so
these are the only quantities we need to evolve in order to model the
full dynamics. All other relevant properties of the population after
crossover can be found from the maximum entropy ansatz.  A more
general method is to follow the evolution of a number of cumulants
explicitly, as in references~\cite{PBS2,Ratt}, but this is unnecessary
here because of the special form of crossover used, which is not
appropriate in problems with stronger spatial interactions.

\subsection{The cumulants and correlation} 

The cumulants of the overlap distribution within the population are
robust statistics which are often reasonably stable to fluctuations
between runs of the GA, so that they average well~\cite{PBS2}. The
first two cumulants are the mean and variance respectively, while the
higher cumulants describe the deviation from a Gaussian
distribution. The third and fourth cumulants are related to the
skewness and kurtosis of the population respectively.  A population
member, labelled $\alpha$, is associated with overlap $R_\alpha$
defined in equation~(\ref{def_R}). The cumulants of the overlap
distribution within a finite population can be generated from the
logarithm of a partition function,
\begin{equation}
        Z = \sum_{\alpha=1}^P \exp(\gamma R_\alpha)
\end{equation}
where $P$ is the population size. If $\kappa_n$ is the $n$th cumulant,
then,
\begin{equation}
        \kappa_n = \lim_{\gamma\rightarrow
        0}\frac{\partial^n}{\partial\gamma^n}\log Z
\end{equation}
The partition function holds all the information required to determine
the cumulants of the distribution of overlaps within the population.

The correlation within the population is a measure of the microscopic
similarity of population members and is important because selection
correlates a finite population, sometimes leading to premature
convergence to poor solutions. It is also important in calculating the
effect of crossover, since this involves the interaction of different
population members and a higher correlation leads to less disruption
on average. The correlation between two population members, $\alpha$
and $\beta$, is $q_{\alpha\beta}$ and is defined by,
\begin{equation}
        q_{\alpha\beta} = \frac{1}{N}\sum_{i=1}^N w_i^\alpha w_i^\beta
\end{equation}
The mean correlation is $q$ and is defined by,
\begin{equation}
        q = \frac{2}{P(P-1)}\sum_{\alpha=1}^P\sum_{\beta > \alpha}
        q_{\alpha\beta}
\end{equation} 

In order to model a finite population we consider that $P$ population
members are randomly sampled from an infinite population, which is
described by a set of infinite population cumulants,
$K_n$~\cite{adam}. The expectation values for the mean correlation and
the first cumulant of a finite population are equal to the infinite
population values. The higher cumulants are reduced by a factor which
depends on the population size, \numparts
\begin{eqnarray}
        \kappa_1 & = & K_1 \label{k1} \\ \kappa_2 & = & P_2 K_2
        \label{k2} \\ \kappa_3 & = & P_3 K_3 \label{k3} \\ \kappa_4 &
        = & P_4 K_4 - 6P_2(K_2)^2/P \label{k4}
\end{eqnarray}
\endnumparts Here, $P_2$, $P_3$ and $P_4$ give finite population
corrections to the infinite population result (see
reference~\cite{PBS2} for a derivation),
\begin{equation}
        P_2 = 1 - \frac{1}{P} \qquad P_3 = 1 - \frac{3}{P} +
\frac{2}{P^2} \qquad P_4 = 1 - \frac{7}{P} + \frac{12}{P^2} -
\frac{6}{P^3}
\end{equation}
Although we model the evolution of a finite population, it is more
natural to follow the macroscopics associated with the infinite
population from which the finite population is
sampled~\cite{adam}. The expected cumulants of a finite population can
be retrieved through equations~(\ref{k1}) to (\ref{k4}).

\section{Crossover and mutation}
\label{sec_mutcross}

The mean effects of standard crossover and mutation on the
distribution of overlaps within the population are equivalent to the
paramagnet results given in~\cite{PBS2}. However, bit-simulated
crossover brings the population straight to the fixed point of
standard crossover, which will be assumed to be a maximum entropy
distribution with the correct mean overlap and correlation, as
described in \ref{max_ent}. To model this form of crossover one only
requires knowledge of these two macroscopics, so these are the only
two quantities we need to evolve under selection and mutation.

The mean overlap and correlation after averaging over all mutations
are, \numparts
\begin{eqnarray}
        K_1^{\rm m} & = (1 - 2p_{\rm m})K_1 \\ q_{\rm m} & = (1 -
        2p_{\rm m})^2q
\end{eqnarray}
\endnumparts where $p_{\rm m}$ is the probability of flipping a bit
under mutation~\cite{PBS2}. The higher cumulants after crossover are
required to determine the effects of selection, discussed in the next
section. The mean overlap and correlation are unchanged by crossover
and the other cumulants can be determined by noting that bit-simulated
crossover completely removes the difference between site averages
within and between different population members. For example, terms
like $\langle w_i^\alpha w_j^\beta\rangle_{i\neq j}$ and $\langle
w_i^\alpha w_j^\alpha \rangle_{i \neq j}$ are equal on average. After
cancelling terms of this form one finds that the first four cumulants
of an infinite population after crossover are, \numparts
\begin{eqnarray} K_1^{\rm c} & = & K_1 \label{k2inf} \\ \vspace{6mm}
        K_2^{\rm c} & = & \frac{1}{N}(1-q) \\ K_3^{\rm c} & = &
        -\frac{2}{N^2} \left(K_1 - \frac{1}{N}\sum_{i=1}^N\langle
        w_i^\alpha \rangle_\alpha^3 \right) \label{k3inf} \\ K_4^{\rm
        c} & = & -\frac{2}{N^3}\left(1 - 4q +
        \frac{3}{N}\sum_{i=1}^N\langle w_i^\alpha \rangle_\alpha^4
        \right) \label{k4inf}
\end{eqnarray}
\endnumparts Here, the brackets denote population averages. The third
and fourth order terms in the expressions for the third and fourth
cumulants are calculated in \ref{max_ent} by making a maximum entropy
ansatz. The expected cumulants of a finite population after crossover
are determined from equations~(\ref{k1}) to (\ref{k4}).

\section{The cumulants after selection}
\label{sec_selcum}

Under selection, $P$ new population members are chosen from the
present population with replacement. Following Pr\"{u}gel-Bennett we
split this operation into two stages~\cite{adam}. First we randomly
sample $P$ population members from an infinite population in order to
create a finite population. Then an infinite population is generated
from this finite population by selection. The proportion of each
population member represented in the infinite population after
selection is equal to its probability of being selected, which is
defined below. The sampling procedure can be averaged out in order to
calculate the expectation values for the cumulants of the overlap
distribution within an infinite population after selection, in terms
of the infinite population cumulants before selection.

The probability of selecting population member $\alpha$ is $p_\alpha$
and for Boltzmann selection one chooses,
\begin{equation}
        p_{\alpha} = \frac{\e^{-\beta E_\alpha}}{\sum^P \e^{-\beta
        E_\alpha}} \label{p_alpha}
\end{equation}
where $\beta$ is the selection strength and the denominator ensures
 that the probability is correctly normalized.  Here, $E_\alpha$ is
 the training energy of population member $\alpha$.

One can then define a partition function for selection,
\begin{equation}
        Z_{\rm s} = \sum_{\alpha=1}^P \exp(-\beta E_\alpha + \gamma
        R_\alpha)
\end{equation} 
The logarithm of this quantity generates the cumulants of the overlap
distribution for an infinite population after selection,
\begin{equation}
        K_n^{\rm s} = \lim_{\gamma\rightarrow 0}
        \frac{\partial^n}{\partial\gamma^n}\log Z_{\rm s}
\end{equation}
One can average this quantity over the population by assuming each
population member is independently selected from an infinite
population with the correct cumulants,
\begin{equation}
        \langle \log Z_{\rm s} \rangle = \left( \prod_{\alpha=1}^P
        \int\!\!\d R_\alpha\,\d E_\alpha\,
        p(R_\alpha)\,p(E_\alpha|R_\alpha) \right) \log Z_{\rm s}
        \label{logZ}
\end{equation}
where $p(E|R)$ determines the stochastic relationship between energy
and overlap as defined in equation~(\ref{def_p}) which will be
approximated by the Gaussian distribution in
equation~(\ref{def_p_g}). Following Pr\"{u}gel-Bennett and Shapiro one
can use Derrida's trick and express the logarithm as an integral in
order to decouple the average~\cite{Derrida,PBS}.
\begin{eqnarray}
        \langle \log Z_{\rm s} \rangle & = & \int_0^\infty\!\! \d t \:
        \frac{\e^{-t} - \langle \e^{-tZ_{\rm s}} \rangle}{t} \nonumber
        \\ & = & \int_0^\infty\!\! \d t \: \frac{\e^{-t} -
        f^P(t,\beta,\gamma)}{t}
\label{log_Z}
\end{eqnarray}
where,
\begin{equation}
        f(t,\beta,\gamma) = \int \!\d R\,\d E\,p(R)\,p(E|R)
        \exp\!\left(-t\e^{-\beta E + \gamma R}\right) \label{ft}
\end{equation}
The distribution of overlaps within an infinite population is
approximated by a cumulant expansion around a Gaussian
distribution~\cite{PBS2},
\begin{equation}
        p(R) = \frac{1}{\sqrt{2\pi
        K_2}}\exp\!\left(\frac{-(R-K_1)^2}{2 K_2}\right)\left[1 +
        \sum_{n=3}^{n_c} \frac{K_n}{K_2^{n/2}} \:
        u_n\!\!\left(\frac{R-K_1}{\sqrt{K_2}}\right)\right]
\label{cum_exp}
\end{equation}
where $u_n(x) = (-1)^n\e^{\frac{x^2}{2}}\frac{\rm d}{\rm d
x^n}\e^{\frac{-x^2}{2}}/n!$ are scaled Hermite polynomials. Four
cumulants were used for the simulations presented in
section~\ref{sec_sim} and the third and fourth Hermite polynomials are
$u_3(x) = (x^3 - 3x)/3!$ and $u_4(x) = (x^4 - 6x^2 + 3)/4!$. This
function is not a well defined probability distribution since it is
not necessarily positive, but it has the correct cumulants and
provides a good approximation. In general, the integrals in
equations~(\ref{log_Z}) and (\ref{ft}) have to be computed
numerically, as was the case for the simulations presented in
section~\ref{sec_sim}.

\subsection{Weak selection and large $N$}
\label{sec_selapp}

It is instructive to expand in small $\beta$ and large $N$, as this
shows the contributions for each cumulant explicitly and gives some
insight into how the size of the training set affects the
dynamics. Since the variance of the population is $O(1/N)$ it is
reasonable to expand the mean of $p(E|R)$, defined in
equation~(\ref{<E|R>}), around the mean of the population in this
limit ($R \simeq K_1$). It is also assumed that the variance of
$p(E|R)$ is well approximated by its leading term and this assumption
may break down if the gradient of the noise becomes important. Under
these simplifying assumptions one finds,
\begin{eqnarray}
        E_{\rm g}(R) \:\simeq \:\frac{\lambda N}{\pi}\left(
        \cos^{-1}(K_1) - \frac{(R - K_1)}{\sqrt{1 - K_1^2}} \right)
        \label{app_mean}\\ \sigma^2 \:\simeq \:\frac{\lambda
        N}{\pi}\cos^{-1}(K_1)\left(1 -
        \frac{1}{\pi}\cos^{-1}(K_1)\right) \label{app_var}
\end{eqnarray}

Following Pr\"{u}gel-Bennett and Shapiro~\cite{PBS}, one can expand
the integrand in equation~(\ref{log_Z}) for small $\beta$ (as long as
$\lambda$ is at least $O(1)$ so that the variance of $p(E|R)$ is
$O(N)$),
\begin{equation}
        f^P(t,\beta,\gamma) \simeq
        \exp(-tP\hat{\rho}_1(\beta,\gamma))\left(1 +
        \frac{Pt^2}{2}\left(\hat{\rho}_2(\beta,\gamma) -
        \hat{\rho}_1^2(\beta,\gamma)\right)\right)
\end{equation}
where,
\begin{equation}
        \hat{\rho}_n(\beta,\gamma) = \int \!\!\d R\,\d E\, p(R) \,
        p(E|R) \:\e^{n(-\beta E + \gamma R)} \label{rhohat}
\end{equation}
We approximate $p(E|R)$ by a Gaussian whose mean and variance given in
equations~(\ref{app_mean}) and (\ref{app_var}). Completing the
integral in equation~(\ref{log_Z}), one finds an expression for the
cumulants of an infinite population after selection,
\begin{equation}
        K_n^{\rm s} = \lim_{\gamma\rightarrow
        0}\frac{\partial^n}{\partial
        \gamma^n}\left[\log(P\rho_1(k\beta,\gamma)) -
        \frac{\e^{(\beta\sigma)^2}}{2P}\left(\frac{\rho_2(k\beta,\gamma)}{\rho_1^2(k\beta,\gamma)}\right)\right]     
\label{kns}
\end{equation}
where,
\begin{eqnarray}
        \rho_n(k\beta,\gamma) & = & \int \!\!\d R\, p(R)\e^{nR(k\beta
        + \gamma)} \nonumber \\ & = & \exp\!\left(\sum_{i=1}^\infty
        \frac{n^i(k\beta+\gamma)^i K_i}{i!}\right) \label{def_rhon}
\end{eqnarray}
Here, a cumulant expansion has been used. The parameter $k$ is the
constant of proportionality relating the generalization error to the
overlap in equation~(\ref{app_mean}) (constant terms are irrelevant,
as Boltzmann selection is invariant under the addition of a constant
to the energy).
\begin{equation}
         k = \frac{\lambda N}{\pi\sqrt{1 - K_1^2}} \label{def_k}
\end{equation}

For the first few cumulants of an infinite population after selection
one finds, \numparts
\begin{eqnarray}
        K_1^{\rm s} & = & K_1 + \left(1 -
        \frac{\e^{(\beta\sigma)^2}}{P}\right)k\beta K_2 + O(\beta^2)\\
        K_2^{\rm s} & = & \left(1 -
        \frac{\e^{(\beta\sigma)^2}}{P}\right)K_2 + \left(1 -
        \frac{3\e^{(\beta\sigma)^2}}{P}\right)k\beta K_3 +
        O(\beta^2)\\ K_3^{\rm s} & = & \left(1 -
        \frac{3\e^{(\beta\sigma)^2}}{P}\right)K_3 -
        \frac{6\e^{(\beta\sigma)^2}}{P}k\beta K_2^2 + O(\beta^2)
        \label{k3s}
\end{eqnarray}
\endnumparts The expected cumulants of a finite population after
selection are retrieved through equations~(\ref{k1}) to (\ref{k4}).
For the zero noise case ($\sigma = 0$) this is equivalent to selecting
directly on overlaps (with energy $-R$), with selection strength
$k\beta$. We will therefore call $k\beta$ the effective selection
strength. It has previously been shown that this parameter should be
scaled inversely with the standard deviation of the population in
order to make continued progress under selection, without converging
too quickly~\cite{PBS2}. Strictly speaking, we can only use
information about the distribution of energies since the overlaps will
not be known in general, but to first order in $R-K_1$ this is
equivalent to scaling the selection strength inversely to the standard
deviation of the energy distribution. As in the problems considered in
reference~\cite{PBS2}, the finite population effects lead to a reduced
variance and an increase in the magnitude of the third cumulant,
related to the skewness of the population. This leads to an
accelerated reduction in variance under further selection. The noise
due to the finite training set increases the size of the finite
population effects. The other genetic operators, especially crossover,
reduce the magnitude of the higher cumulants to allow further progress
under selection.
 
\section{The correlation after selection}
\label{sec_cor}

To model the full dynamics, it is necessary to evolve the mean
correlation within the population under selection. This is rather
tricky, as it requires knowledge of the relationship between overlaps
and correlations within the population. To make the problem tractable,
it is assumed that before selection the population is at maximum
entropy with constraints on the mean overlap and correlation within
the population, as discussed in \ref{max_ent}. The calculation
presented here is similar to that presented elsewhere~\cite{Ratt},
except for a minor refinement which seems to be important when
considering problems with noise under selection.

The correlation of an infinite population after selection from a
finite population is given by,
\begin{eqnarray}
        q_{\rm s} & = & \sum_{\alpha=1}^P p_\alpha^2(1 -
        q_{\alpha\alpha}) + \sum_{\alpha=1}^P\sum_{\beta=1}^P p_\alpha
        p_\beta q_{\alpha\beta} \nonumber \\ & = & \Delta q_\d \: + \:
        q_\infty \label{q_s}
\end{eqnarray}
where $p_\alpha$ is the probability of selection, defined in
equation~(\ref{p_alpha}). The first term is due to the duplication of
population members under selection, while the second term is due to
the natural increase in correlation as the population moves into a
region of lower entropy. The second term gives the increase in the
correlation in the infinite population limit, where the duplication
term becomes negligible. An extra set of variables $q_{\alpha\alpha}$
are assumed to come from the same statistics as the distribution of
correlations within the population. Recall that the expectation value
for the correlation of a finite population is equal to the correlation
of the infinite parent population from which it is sampled.

\subsection{Natural increase term}

We estimate the conditional probability distribution for correlations
given overlaps before selection $p(q_{\alpha\beta}|R_\alpha,R_\beta)$
by assuming the weights within the population are distributed
according to the maximum entropy distribution described in
\ref{max_ent}. Then $q_\infty$ is simply the correlation averaged over
this distribution and the distribution of overlaps after selection,
$p_{\rm s}(R)$.
\begin{equation}
        q_\infty = \int\!\! \d q_{\alpha\beta} \, \d R_\alpha \, \d
        R_\beta \, p_{\rm s}(R_\alpha) p_{\rm s}(R_\beta)
        p(q_{\alpha\beta}|R_\alpha,R_\beta)\,q_{\alpha\beta}
        \label{q_infty} \\
\end{equation}
This integral can be calculated for large $N$ by the saddle point
method and we find that in this limit the result only depends on the
mean overlap after selection (see \ref{app_cond}).
\begin{equation}
        q_\infty(y) = \frac{1}{N}\sum_{i=1}^N \left(\ \frac{W_i +
        \tanh(y)}{1 + W_i\tanh(y)} \right)^2 \label{qqs}
\end{equation}
where,
\begin{equation}
        K_1^{\rm s} = \frac{1}{N}\sum_{i=1}^N \frac{W_i + \tanh(y)}{1
        + W_i\tanh(y)} \label{qk1s}
\end{equation}
The natural increase contribution to the correlation $q_\infty$ is an
implicit function of $K_1^{\rm s}$ through $y$, which is related to
$K_1^{\rm s}$ by equation~(\ref{qk1s}). Here, $W_i$ is the mean weight
at site $i$ before selection (recall that we have chosen the teacher's
weights to be $t_i = 1$ at every site, without loss of generality) and
for a distribution at maximum entropy one has,
\begin{equation}
        W_i = \tanh(z + x\eta_i) \\
\end{equation}
The Lagrange multipliers, $z$ and $x$, are chosen to enforce
constraints on the mean overlap and correlation within the population
before selection and $\eta_i$ is drawn from a Gaussian distribution
with zero mean and unit variance (see \ref{max_ent}).

It is instructive to expand in $y$, which is appropriate in the weak
selection limit. In this case one finds,
\begin{eqnarray}
        K_1^{\rm s} = K_1^{\rm c} + y(N K_2^{\rm c}) +
        \frac{y^2}{2}(N^2 K_3^{\rm c}) + \cdots \\ q_\infty(y) = q -
        y(N^2 K_3^{\rm c}) - \frac{y^2}{2}(N^3 K_4^{\rm c}) + \cdots
        \label{q_inf}
\end{eqnarray}
where $K_n^{\rm c}$ are the infinite population expressions for the
cumulants after bit-simulated crossover, when the population is
assumed to be at maximum entropy (defined in equations (\ref{k2inf})
to (\ref{k4inf}) up to the fourth cumulant). Here, $y$ plays the role
of the effective selection strength in the associated infinite
population problem, so for an infinite population one could simply set
$y = k\beta/N$, where $k$ is defined in equation~(\ref{def_k}). To
calculate the correlation after selection, we solve
equation~(\ref{qk1s}) for $y$ and then substitute this value into the
equation~(\ref{qqs}) to calculate $q_\infty$. In general this must be
done numerically, although the weak selection expansion can be used to
obtain an analytical result which gives a very good approximation in
many cases. Notice that the third cumulant in equation~(\ref{q_inf})
will be negative for $K_1>0$ because of the negative entropy gradient
and this will accelerate the increased correlation under selection.

\subsection{Duplication term}
\label{sec_dup}

The duplication term $\Delta q_\d$ is defined in
equation~(\ref{q_s}). As in the partition function calculation
presented in section~\ref{sec_selcum}, population members are
independently averaged over a distribution with the correct cumulants,
\begin{eqnarray}
\fl \Delta q_\d = P\left( \prod_{\alpha=1}^P \int \!\!\d R_\alpha \,
\d E_\alpha \,\d q_{\alpha\alpha}p(R_\alpha) \, p(E_\alpha|R_\alpha)
\, p(q_{\alpha\alpha}|R_\alpha,R_\alpha) \right) \frac{(1 -
q_{\alpha\alpha})\e^{-2\beta E_\alpha}}{(\sum_\alpha \e^{-\beta
E_\alpha})^2} \nonumber \\ \lo = P\left( \prod_{\alpha=1}^P \int
\!\!\d R_\alpha \cdots \right)(1 - q_{\alpha\alpha})\exp(-2\beta
E_\alpha) \int_0^\infty \!\!\!\d t\,t\,\exp\left(-t\sum_\alpha
\e^{-\beta E_\alpha}\right)
\end{eqnarray}
Here, $q_{\alpha\alpha}$ is a construct which comes from the same
statistics as the correlations between distinct population members.
The integral in $t$ removes the square in the denominator and
decouples the average,
\begin{equation}
        \Delta q_\d \: = \: P\!\!\int_0^\infty \!\!\!\d t\,t\,
        f(t)\,g^{P-1}(t) \label{q_d}
\end{equation}
where,
\begin{eqnarray}
        f(t) & = & \int \!\!\d R \, \d E \, \d q\, p(R) \, p(E|R) \,
        p(q|R,R) \:(1 - q)\exp(-2\beta E - t\e^{-\beta E}) \\ g(t) & =
        & \int \!\!d R \, \d E \, p(R) p(E|R) \exp(-t\e^{-\beta E})
\end{eqnarray}
The overlap distribution $p(R)$ will be approximated by the cumulant
expansion in equation~(\ref{cum_exp}) and $p(q|R,R)$ by the
distribution derived in \ref{app_cond}. In general, it would be
necessary to calculate these integrals numerically, but the
correlation distribution is difficult to deal with as it requires the
numerical reversion of a saddle point equation.

Instead, we expand for small $\beta$ and large $N$ as we did for the
selection calculation in section~\ref{sec_selapp} (this approximation
is only used for the term involving the correlation in
equation~(\ref{q_d}) for the simulations presented in
section~\ref{sec_sim}). In this case one finds,
\begin{eqnarray}
        f(t)\,g^{P-1}(t) \; & \simeq &\;
                        \hat{\rho}(2\beta)\exp\!\left[-t\left(
                        (P-1)\hat{\rho}(\beta) +
                        \frac{\hat{\rho}(3\beta)}{\hat{\rho}(2\beta)}
                        \right)\right] \nonumber \\ & & -
                        \hat{\rho}_q(2\beta)\exp\!\left[-t\left(
                        (P-1)\hat{\rho}(\beta) +
                        \frac{\hat{\rho}_q(3\beta)}{\hat{\rho}_q(2\beta)}
                        \right)\right]
\end{eqnarray}
where,
\begin{eqnarray}
        \hat{\rho}(\beta) & = & \int \!\!\d R\,\d E\, p(R) \, p(E|R)
        \:\e^{-\beta E} \\ \hat{\rho}_q(\beta) & = & \int \!\!\d R\,\d
        E\, p(R) \, p(E|R) \int\!\!\d q \: p(q|R,R) \,q\,\e^{-\beta E}
\end{eqnarray}  
Completing the integral in equation~(\ref{q_d}) one finds,
\begin{equation}
        \Delta q_\d = \frac{\hat{\rho}(2\beta) -
        \hat{\rho}_q(2\beta)}{P\hat{\rho}^2(\beta)} +
        O\!\!\left(\frac{1}{P^2}\right)
\end{equation}
We express $\hat{\rho}_q(\beta)$ in terms of the Fourier transform of
the distribution of correlations, which is defined in
equation~(\ref{ftq}),
\begin{equation}
        \hat{\rho}_q(\beta) = \lim_{t \rightarrow
        0}\frac{\partial}{\partial t} \log\!\left(\int \!\!\d R\,\d
        E\, p(R) \, p(E|R) \hat{\rho}(-\i t|R,R) \,\e^{-\beta
        E}\right)\!\hat{\rho}(\beta)
\end{equation}
The integrals can be calculated by expressing $p(E|R)$ by the same
approximate form as in section~\ref{sec_selapp} and using the saddle
point method to integrate over the Fourier transform as in
\ref{app_cond}.

Eventually one finds,
\begin{equation}
        \Delta q_\d \; = \; \frac{\e^{(\beta\sigma)^2} [1 -
        q_\infty(2k\beta/N)] \rho_2(k\beta,0)}{P \rho_1^2(k\beta,0) }
        \; + \; O\!\!\left(\frac{1}{P^2}\right) \label{dqd}
\end{equation}
where $q_\infty(y)$ is defined in equation~(\ref{qqs}) and
$\rho_n(k\beta,\gamma)$ is defined in equation~(\ref{def_rhon}).

It is instructive to expand in $\beta$ as this shows the contributions
from each cumulant explicitly.  To do this we use the cumulant
expansion described in equation~(\ref{cum_exp}) and to third order in
$\beta$ for three cumulants one finds,
\begin{equation}
\fl \Delta q_\d \; \simeq \; \frac{\e^{(\beta\sigma)^2}}{P} \left[1 -
q_\infty(2k\beta/N)\right]\left( 1 + K_2(k\beta)^2 - K_3(k\beta)^3 +
O(\beta^4) \right)
\end{equation}
The $q_\infty$ term has not been expanded out since it contributes
terms of $O(1/N)$ less than these contributions for each
cumulant. Selection leads to a negative third cumulant (see
equation~(\ref{k3s})), which in turn leads to an accelerated increase
in correlation under further selection. Crossover reduces this effect
by reducing the magnitude of the higher cumulants.

\section{Dynamic population resizing}
\label{sec_rescale}

The noise introduced by the finite sized training set increases the
magnitude of the detrimental finite population terms in selection. In
the limit of weak selection and large problem size discussed in
sections~\ref{sec_selapp} and \ref{sec_dup}, this can be compensated
for by increasing the population size.  The terms which involve noise
in equations (\ref{kns}) and (\ref{dqd}) can be removed by an
appropriate population resizing,
\begin{equation}
        P = P_0\exp[(\beta\sigma)^2]
\end{equation}
Here, $P_0$ is the population size in the infinite training set, zero
noise limit. Since these are the only terms in the expressions
describing the dynamics which involve the finite population size, this
effectively maps the full dynamics onto the infinite training set
case.

For zero noise the selection strength should be scaled so that the
effective selection strength $k\beta$ is inversely proportional to the
standard deviation of the population~\cite{PBS},
\begin{equation}
        \beta = \frac{\beta_{\rm s}}{k\sqrt{\kappa_2}}
\end{equation}
Here, $k$ is defined in equation~(\ref{def_k}) and $\beta_{\rm s}$ is
the scaled selection strength and remains fixed throughout the
search. Recall that $\kappa_2$ is the expected variance of a finite
population, which is related to the variance of an infinite population
through equation~(\ref{k2}). One could also include a factor of
$\sqrt{\log P}$ to compensate for changes in population size, as in
reference~\cite{PBS2}, but this term is neglected here. The resized
population is then,
\begin{eqnarray}
        P & = & P_0\exp\left(\frac{(\beta_{\rm s}\sigma)^2}{k^2
        \kappa_2}\right) \nonumber \\ & = &
        P_0\exp\left(\frac{\beta_{\rm s}^2 (1 -
        \kappa_1^2)\cos^{-1}(\kappa_1)(\pi -
        \cos^{-1}(\kappa_1))}{\lambda N \kappa_2}\right)
\label{rescale}
\end{eqnarray}
Notice that the exponent in this expression is $O(1)$, so this
population resizing does not blow up with increasing problem size. One
might therefore expect this problem to scale with $N$ in the same
manner as the zero-noise, infinite training set case, as long as the
batch size is $O(N)$.

Baum \etal have shown that a closely related GA scales as
$O(N\log_2^2N)$ on this problem if the population size is sufficiently
large so that alleles can be assumed to come from a binomial
distribution~\cite{Baum}. This is effectively a maximum entropy
assumption with a constraint on the mean overlap alone. They use
culling selection, where the best half of the population survives each
generation leading to a change in the mean overlap proportional to the
population's standard deviation. Our selection scaling also leads to a
change in the mean of this order and the algorithms may therefore be
expected to compare closely. The expressions derived here do not rely
on a large population size and are therefore more general.

In the infinite population limit it is reasonable to assume $N\kappa_2
\simeq 1 - \kappa_1^2$ which is the relationship between mean and
variance for a binomial distribution, since in this limit the
correlation of the population will not increases due to duplication
under selection.  In this case the above scaling results in a
monotonic decrease in population size, as $\kappa_1$ increases over
time. This is easy to implement by removing the appropriate number of
population members before each selection.

In a finite population the population becomes correlated under
selection and the variance of the population is usually less than the
value predicted by a binomial distribution.  In this case the
population size may have to be increased, which could be implemented
by producing a larger population after selection or crossover. This is
problematic, however, since increasing the population size leads to an
increase in the correlation and a corresponding reduced
performance. In this case the dynamics will no longer be equivalent to
the infinite training set situation.

Instead of varying the population size, one can fix the population
size and vary the size of the training batches. In this case one
finds,
\begin{equation}
        \lambda = \frac{\beta_{\rm
        s}^2(1-\kappa_1^2)\cos^{-1}(\kappa_1)(\pi -
        \cos^{-1}(\kappa_1))}{N\kappa_2\log(P/P_0)}
        \label{scale_alpha}
\end{equation}

\begin{figure}[h]
        \setlength{\unitlength}{1.0cm}
        \begin{center}
        \begin{picture}(8,6)
        \put(0,0){\epsfig{figure=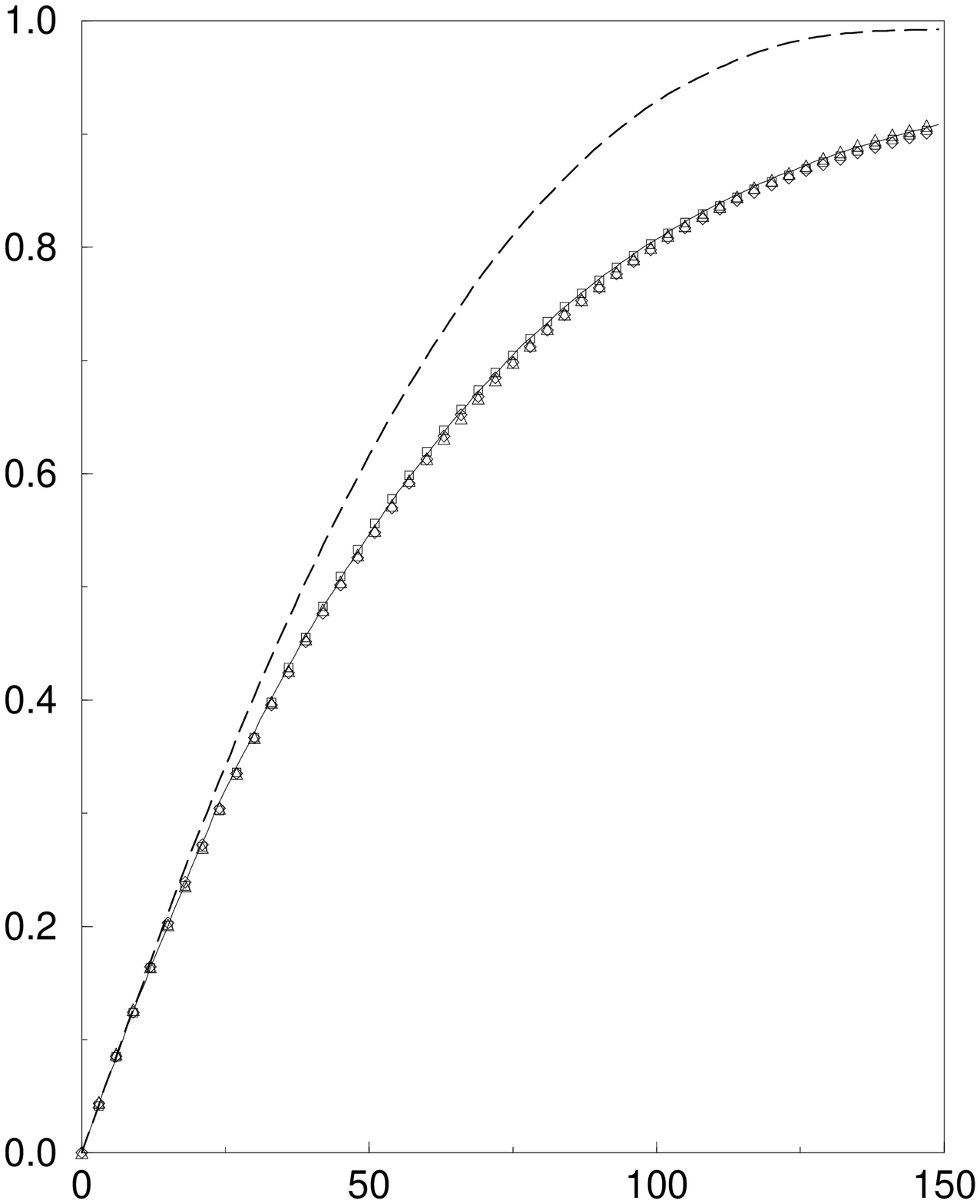,width=8.0cm,height=6.0cm}}
        \put(3.7,0.6){\epsfig{figure=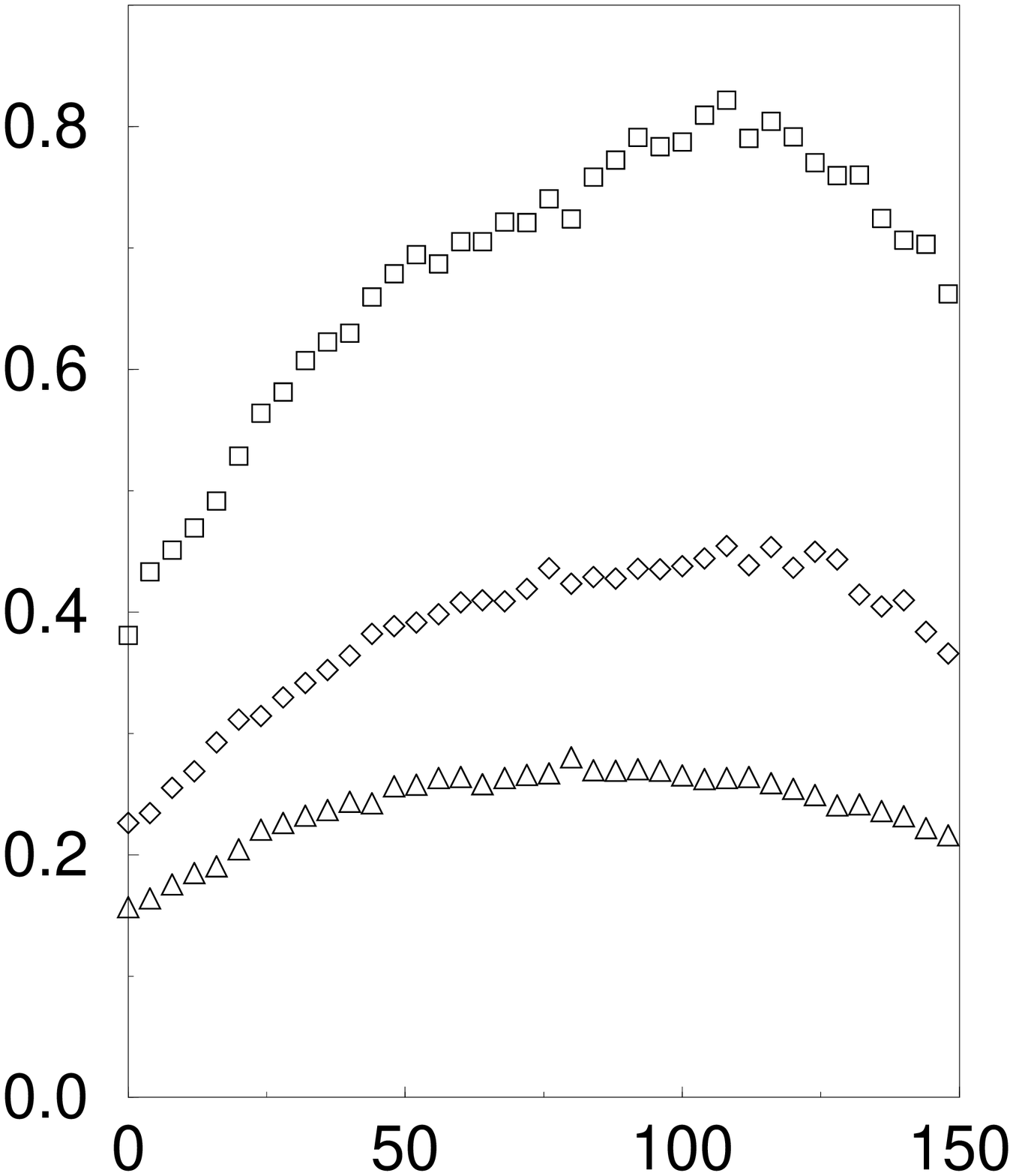,width=4.0cm,height=2.7cm}}
        \put(3.4,-0.6){\mbox{Generation}}
        \put(3.5,2.1){\mbox{\small{$\lambda$}}}
        \put(-0.5,4.3){\mbox{$\kappa_1$}}
        \end{picture}
        \end{center}
\caption{The mean overlap between teacher and pupil within the population 
is shown each generation for a GA training a binary perceptron to
generalize from examples produced by a teacher perceptron. The results
were averaged over $100$ runs and training batch sizes were chosen
according to equation~(\ref{scale_alpha}), leading to the trajectories
converging onto the infinite training set result where $E=E_{\rm
g}(R)$.  The solid curve is for the infinite training set with
$P_0=60$ and the finite training set results are for $P = 90$
(\opensqr), $120 (\diamond)$ and $163 (\triangle)$. Inset is the mean
choice of $\lambda$ each generation. The dashed line is the infinite
training set result for $P = 163$, showing that there is significant
potential variability of trajectories under different batch sizing
schemes. The other parameters were $N=279$, $\beta_{\rm s}=0.25$ and
$p_{\rm m}=0.001$.}
        \label{fig_scale}
\end{figure}

Figure~\ref{fig_scale} shows how choosing the batch size each
generation according to equation~(\ref{scale_alpha}) leads to the
dynamics converging onto the infinite training set dynamics where the
training energy is equal to the generalization error. The infinite
training set result for the largest population size is also shown, as
this gives some measure of the potential variability of trajectories
available under different batch sizing schemes.  Any deviation from
the weak selection, large $N$ limit is not apparent here. To a good
approximation it seems that the population resizing in
equation~(\ref{rescale}) and the corresponding batch sizing expression
in equation~(\ref{scale_alpha}) are accurate, at least as long as
$\lambda$ is not too small.

\subsection{Optimal batch size}

In the previous section it was shown how the population size could be
changed to remove the effects of noise associated with a finite
training set. If we use this population resizing then it is possible
to define an optimal size of training set, in order to minimize the
computational cost of energy evaluation. This choice will also
minimize the total number of training examples presented when
independent batches are used. This may be expected to provide a useful
estimate of the appropriate sizing of batches in more efficient
schemes, where examples are recycled, as long as the total number of
examples used significantly exceeds the threshold above which
over-training is impossible.

We assume that computation is mainly due to energy evaluation and note
that there are $P$ energy evaluations each generation with computation
time for each scaling as $\lambda$. If the population size each
generation is chosen by equation~(\ref{rescale}), then the computation
time $\tau_c$ (in arbitrary units) is given by,
\begin{equation}
        \tau_c \: = \: \lambda
\,\exp\!\left(\frac{\lambda_o}{\lambda}\right) \qquad \lambda_o =
\frac{\beta_{\rm s}^2(1-\kappa_1^2)\cos^{-1}(\kappa_1)(\pi -
\cos^{-1}(\kappa_1))}{N\kappa_2}
\end{equation}
The optimal choice of $\lambda$ is given by the minimum of $\tau_c$,
which is at $\lambda_o$.  Choosing this batch size leads to the
population size being constant over the whole GA run and for optimal
performance one should choose,
\begin{eqnarray}
        P & = & P_0 \, \e^1 \: \simeq \: 2.73 P_0 \\ \lambda & = &
        \lambda_o
\end{eqnarray}
where $P_0$ is the population size used for the zero noise, infinite
training set GA. Notice that it is not necessary to determine $P_0$ in
order to choose the size of each batch, since $\lambda_o$ is not a
function of $P_0$. Since the batch size can now be determined
automatically, this reduces the size of the GA's parameter space
significantly.
 
One of the runs in figure~\ref{fig_scale} is for this choice of $P$
and $\lambda$, showing close agreement to the infinite training set
dynamics ($P = 163 \, \simeq \, P_0\e$). In general, the first two
cumulants change in a non-trivial manner each generation and their
evolution can be determined by simulating the dynamics, as described
in section~\ref{sec_sim}.

\section{Simulating the dynamics}
\label{sec_sim}

In sections~\ref{sec_mutcross}, \ref{sec_selcum} and \ref{sec_cor},
difference equations were derived for the mean effect of each operator
on the mean overlap and correlation within the population. The full
dynamics of the GA can be simulated by iterating these equations
starting from their initial values, which are zero. The equations for
selection also require knowledge of the higher cumulants before
selection, which are calculated by assuming a maximum entropy
distribution with constraints on the two known macroscopics (see
equations (\ref{k2inf}) to (\ref{k4inf})).  We used four cumulants and
the selection expressions were calculated numerically, although for
weak selection the analytical results in section~(\ref{sec_selapp})
were also found to be very accurate. The largest overlap within the
population was estimated by assuming population members were randomly
selected from a distribution with the correct
cumulants~\cite{PBS2}. This assumption breaks down towards the end of
the search, when the population is highly correlated and the higher
cumulants become large, so that four cumulants may not describe the
population sufficiently well.

\begin{figure}[h]
        \setlength{\unitlength}{1.0cm}
        \begin{center}
        \begin{picture}(8,6)
        \put(0,0){\epsfig{figure=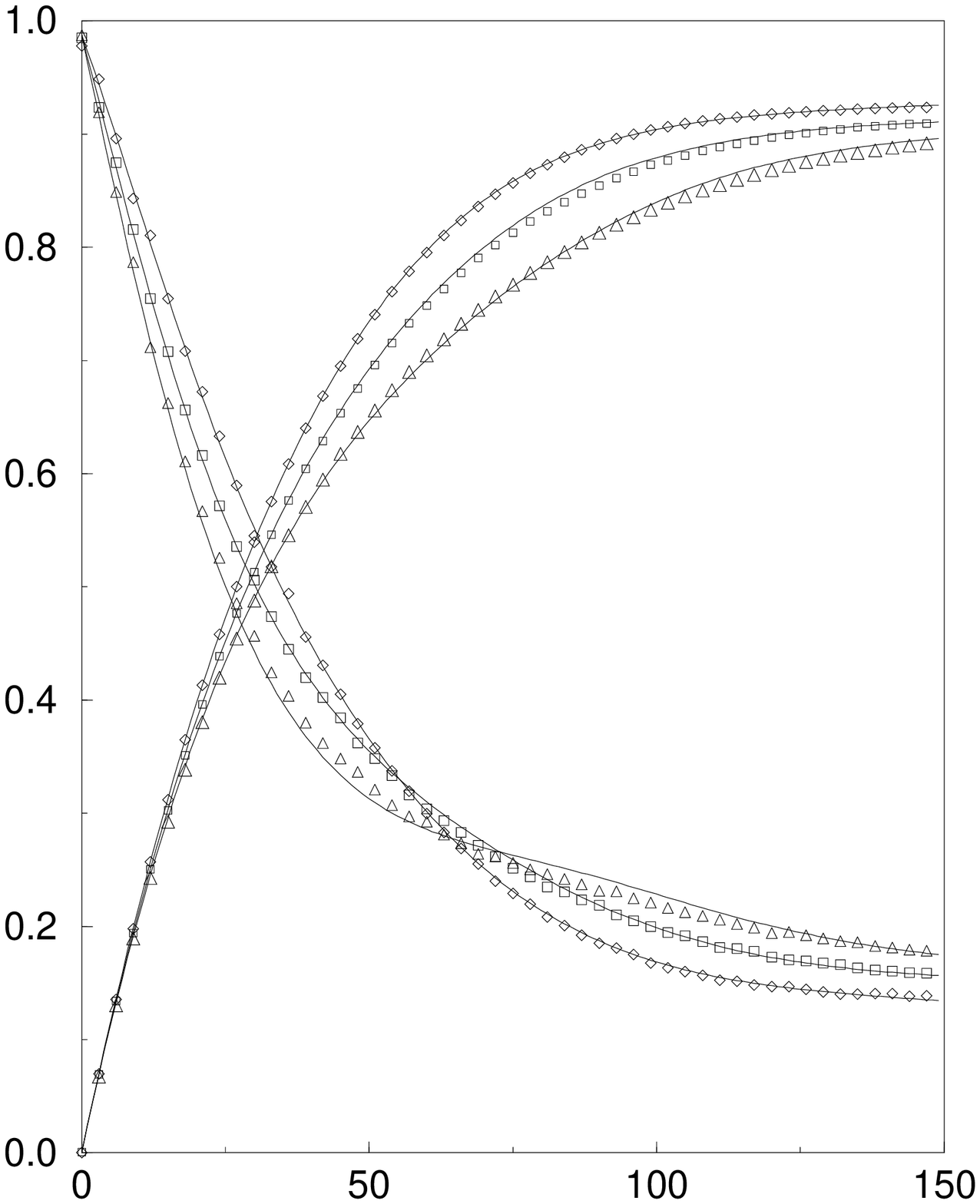,width=8.0cm,height=6.0cm}}
        \put(3.4,-0.6){\mbox{Generation}}
        \put(3.5,4.0){\mbox{$\kappa_1$}}
        \put(3.5,2.3){\mbox{$N\kappa_2$}}
        \end{picture}
        \end{center}
\caption{The theory is compared to averaged results from a GA training 
a binary perceptron to generalize from examples produced by a teacher
perceptron.  The mean and variance of the overlap distribution within
the population are shown, averaged over $1000$ runs, with the solid
lines showing the theoretical predictions. The infinite training set
result~($\Diamond$) is compared to results for a finite training set
with $\lambda = 0.65$~(\opensqr) and $\lambda =
0.39$~($\triangle$). The other parameters were $N=155$, $\beta_{\rm
s}=0.3$, $p_{\rm m}=0.005$ and the population size was $80$.}
        \label{fig_conv}
\end{figure}

\begin{figure}[h]
        \setlength{\unitlength}{1.0cm}
        \begin{center}
        \begin{picture}(8,6)
        \put(0,0){\epsfig{figure=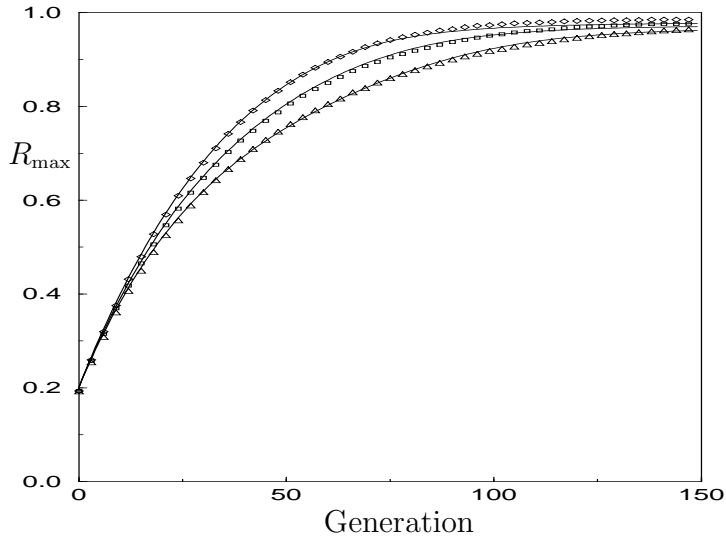,width=8.0cm,height=6.0cm}}
        \put(3.4,-0.6){\mbox{Generation}}
        \put(-0.8,4.3){\mbox{$R_{\mbox{\tiny{max}}}$}}
        \end{picture}
        \end{center}
\caption{The maximum overlap between teacher and pupil is shown each generation,
averaged over the same runs as the results presented in
figure~\ref{fig_conv}.  The solid lines show the theoretical
predictions and the symbols are as in figure~\ref{fig_conv}.}
        \label{fig_best}
\end{figure}

\begin{figure}[h]
        \setlength{\unitlength}{1.0cm}
        \begin{center}
        \begin{picture}(8,6)
        \put(0,0){\epsfig{figure=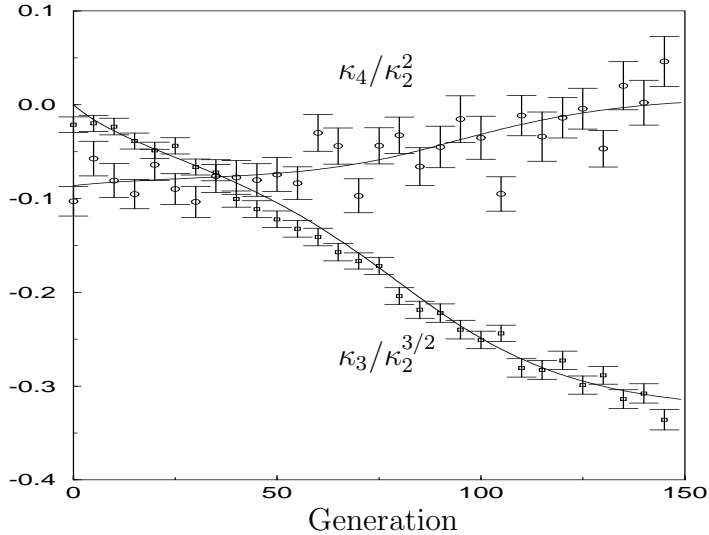,width=8.0cm,height=6.0cm}}
        \put(3.4,-0.6){\mbox{Generation}}
        \put(3.8,5.4){\mbox{$\kappa_4/\kappa_2^2$}}
        \put(3.8,1.6){\mbox{$\kappa_3/\kappa_2^{3/2}$}}
        \end{picture}
        \end{center}
\caption{The skewness and kurtosis of the overlap distribution are shown averaged over the same runs 
as the results presented in figure~\ref{fig_conv} for $\lambda =
0.65$. Averages were taken over cumulants, rather than the ratios
shown. The solid lines show the theoretical predictions for mean
behaviour.}
        \label{fig_conv34}
\end{figure}

Figures~\ref{fig_conv} and \ref{fig_best} show the mean, variance and
largest overlap within the population each generation, averaged over
1000 runs of a GA and compared to the theory. The infinite training
set case, where the training energy is the generalization error, is
compared to results for two values of $\lambda$, showing how
performance degrades as the batch size is reduced. Recall that
$\lambda N$ new patterns are shown to each population member, each
generation, so that the total number of patterns used is $\lambda
N\!\times\! PG$, where $P$ is population size and $G$ is the total
number of generations. The skewness and kurtosis are presented in
figure~\ref{fig_conv34} for one value of $\lambda$, showing that
although there are larger fluctuations in the higher cumulants they
seem to agree sufficiently well to the theory on average. It would
probably be possible to model the dynamics accurately with only three
cumulants, since the kurtosis does not seem to be particularly
significant in these simulations.

These results show excellent agreement with the theory, although there
is a slight underestimate in the best population member for the
reasons discussed above. This is typical of the theory, which has to
be very accurate in order to pick up the subtle effects of noise due
to the finite batch size. Unfortunately, the agreement is less
accurate for low values of $\lambda$, where the noise is
stronger. This may be due to two simplifications. Firstly, we use a
Gaussian approximation for the noise which relies on $\lambda$ being
at least $O(1)$. This could be remedied by expanding the noise in
terms of more than two cumulants as we have done for the overlap
distribution. Secondly, the duplication term in section~\ref{sec_dup}
uses the large $N$, weak selection approximation which also relies on
$\lambda$ being $O(1)$. The error due to this approximation is
minimized by only using the approximation for the term involving the
correlation in equation (\ref{q_d}), with the other term calculated
numerically. It is expected that good results for smaller values of
$\lambda$ would be possible for larger values of $N$, where the
correlation calculation would be more exact.

\section{Conclusion}

A statistical mechanics formalism has been used to solve the dynamics
of a GA for a simple problem from learning theory, generalization in a
perceptron with binary weights.  To make the dynamics tractable, the
case where a new batch of examples was presented to each population
member each generation was considered.  For $O(N)$ training examples
per batch the training energy was well approximated by a Gaussian
distribution whose mean is the generalization error and whose variance
increases as the batch size is reduced.  The use of bit-simulated
crossover, which takes the population straight to the fixed point of
standard crossover, allowed the dynamics to be modelled in terms of
only two macroscopics; the mean correlation and overlap within the
population. The higher cumulants of the overlap distribution after
crossover were required to calculate the effect of selection and were
estimated by assuming maximum entropy with respect to the two known
macroscopics. By iterating difference equations describing the average
effect of each operator on the mean correlation and overlap the
dynamics of the GA were simulated, showing very close agreement with
averaged results from a GA.

Although the difference equations describing the effect of each
operator required numerical enumeration in some cases, analytical
results were derived for the weak selection, large $N$ limit. It was
shown that in this limit a dynamical resizing of the population maps
the finite training set dynamics onto the infinite training set
situation. Using this resizing it is possible to calculate the most
computationally efficient size of population and training batch, since
there is a diminishing return in improved performance as batch size is
increased. For the case of independent training examples considered
here this choice also gives the minimum total number of examples
presented.

In future work it would be essential to look at the situation where
the patterns are recycled, leading to a much more efficient use of
training examples and the possibility of over-training. In this case,
the distribution of overlaps between teacher and pupil would not be
sufficient to describe the population, since the training energy would
then be dependent on the training set. One would therefore have to
include information specific to the training set, such as the mean
pattern per site within the training set. This might be treated as a
quenched field at each site, although it is not obvious how one could
best incorporate such a field into the dynamics.

Another interesting extension of the present study would be to
consider multi-layer networks, which would present a much richer
dynamical behaviour than the single-layer perceptron considered
here. This would bring the formalism much closer to problems of
realistic difficulty. In order to describe the population in this case
it would be necessary to consider the joint distribution of many order
parameters within the population. It would be interesting to see how
the dynamics of the GA compares to gradient methods in networks with
continuous weights, for which the dynamics of generalization for a
class of multi-layer architectures have recently been solved
analytically in the case of on-line learning~\cite{Saad}. In order to
generalize in multi-layer networks it is necessary for the search to
break symmetry in weight space and it would be of great interest to
understand how this might occur in a population of solutions, whether
it would occur spontaneously over the whole population in analogy to a
phase transition or whether components would be formed within the
population, each exhibiting a different broken symmetry. This would
again require the accurate characterization of finite population
effects, since an infinite population might allow the coexistence of
all possible broken symmetries, which is presumably an unrealizable
situation in finite populations.

\ack We would like to thank Adam Pr\"{u}gel-Bennett for many helpful
discussions and for providing code for some of the numerical work used
here. We would also like to thank the anonymous reviewers for making a
number of useful suggestions.  MR was supported by an EPSRC award
(ref. 93315524).

\appendix

\section{The maximum entropy distribution} 
\label{max_ent}

After bit-simulated crossover the population is assumed to be at
maximum entropy with constraints on the mean overlap and correlation
within the population.  This is a special case of the result derived
for the paramagnet by Pr\"{u}gel-Bennett and Shapiro~\cite{PBS2} and
this discussion follows theirs closely.

Let $W_i$ be the mean weight at site $i$ within the population,
\begin{equation}
        W_i = \langle w_i^\alpha \rangle_\alpha =
        \frac{1}{P}\sum_{\alpha=1}^P w_i^\alpha
\end{equation}
To calculate the distribution of this quantity over sites one imposes
constraints on the mean overlap and correlation with Lagrange
multipliers $x$ and $z$,
\begin{eqnarray}
        zPK_1 \: = \: \frac{z}{N}\sum_{\alpha=1}^P\sum_{i=1}^N
w_i^\alpha \: = \: \frac{zP}{N}\sum_{i=1}^N W_i \\ \frac{(xP)^2}{2} q
\: = \: \frac{x^2}{2N}\sum_{\alpha=1}^P\sum_{\beta=1} ^P\sum_{i=1}^N
w_i^\alpha w_i^\beta \: = \: \frac{(xP)^2}{2N}\sum_{i=1}^N W_i^2
\end{eqnarray}
Recall that we have chosen $t_i=1$ at each site without loss of
generality. The correlation expression is for large $P$ and finite
population corrections can be included retrospectively.

Without constraints, the fraction of positive weights at site $i$ is
given by a binomial coefficient,
\begin{equation}
        \Omega(W_i) = \frac{1}{2^P}\left( \begin{array}{c} P \\
                                P(1+W_i)/2
                        \end{array} \right)
\end{equation}
So one can define an entropy,
\begin{eqnarray}
        S(W_i) & = & \log[\Omega(W_i)] \nonumber \\ & \sim &
        -\frac{P}{2}\log(1 - W_i^2) +
        \frac{PW_i}{2}\log\left(\frac{1-W_i}{1+W_i}\right)
\end{eqnarray}
where Stirling's approximation has been used. One can then define a
probability distribution for the $\{W_i\}$ configuration which
decouples at each site,
\begin{eqnarray}
        p(\{W_i\}) \: = \: \prod_{i=1}^N p(W_i) \: = \: \prod_{i=1}^N
        \exp[S(W_i) + zPW_i + (xPW_i)^2/2] \\ p(W_i) \: = \: \int
        \!\!\frac{\d \eta_i}{\sqrt{2\pi}}\,
        \exp\!\left(\frac{-\eta^2_i}{2} + PG(W_i,\eta_i)\right)
\end{eqnarray}
where
\begin{equation}
        G(W_i,\eta_i) = S(W_i)/P + zW_i + x\eta_iW_i
\end{equation}
The maximal value of $G$ with respect to $W_i$ gives the maximum
entropy distribution for $W_i$ at each site. This leads to the
expression,
\begin{equation}
        W_i = \tanh(z + x\eta_i) \label{W_i}
\end{equation}
where $\eta_i$ is drawn from a Gaussian with zero mean and unit
variance.  The constraints can be used to obtain values for the
Lagrange multipliers,
\begin{eqnarray}
        K_1 & = & \frac{1}{N}\sum_{i=1}^N \overline{\tanh(z +
        x\eta_i)} \\ q & = & \frac{1}{N}\sum_{i=1}^N
        \overline{\tanh^2(z + x\eta_i)}
\end{eqnarray}
The bars denote averages over the Gaussian noise which in general must
be done numerically.

The third and fourth order terms in equations (\ref{k3inf}) and
(\ref{k4inf}) can be found once the Lagrange multipliers have been
determined,
\begin{eqnarray}
        \frac{1}{N}\sum_{i=1}^N \langle w_i^\alpha \rangle_{\alpha}^3
        = \overline{\tanh^3(z + x\eta)} \\ \frac{1}{N}\sum_{i=1}^N
        \langle w_i^\alpha \rangle_{\alpha}^4 = \overline{\tanh^4(z +
        x\eta)}
\end{eqnarray}
Again, the bars denote averages over the Gaussian noise.
        
\section{The distribution of correlations}
\label{app_cond}

Rewriting equation~(\ref{q_infty}) we have,
\begin{eqnarray}
        q_\infty & = & \int\!\! \d q_{\alpha\beta} \, \d R_\alpha \,
        \d R_\beta \, p_{\rm s}(R_\alpha) p_{\rm s}(R_\beta) \,
        p(q_{\alpha\beta}|R_\alpha,R_\beta)\, q_{\alpha\beta}
        \nonumber \\ & = & \lim_{t\rightarrow 0}
        \frac{\partial}{\partial t} \log\left( \int \!\!\d R_\alpha
        \,\d R_\beta\: p_{\rm s}(R_\alpha) \, p_{\rm s}(R_\beta) \,
        \hat{\rho}(-\i t|R_\alpha,R_\beta)\right)
\end{eqnarray}
where $\hat{\rho}(-\i t|R_\alpha,R_\beta)$ is the Fourier transform of
$p(q_{\alpha\beta}|R_\alpha,R_\beta)$,
\begin{equation}
        \hat{\rho}(-\i t|R_\alpha,R_\beta) \: = \: \int\!\!\d
        q_{\alpha\beta}
        \:p(q_{\alpha\beta}|R_\alpha,R_\beta)\e^{tq_{\alpha\beta}}
        \label{ftq}
\end{equation}
The conditional probability for correlations
$p(q_{\alpha\beta}|R_\alpha,R_\beta)$ can be defined if weights are
assumed to come from the maximum entropy distribution defined in
\ref{max_ent}. In this case one has,
\begin{eqnarray}
        p(q_{\alpha\beta}|R_\alpha,R_\beta) =
         \frac{p(q_{\alpha\beta},R_\alpha,R_\beta)}{p(R_\alpha,R_\beta)}
         \nonumber \\ = \frac{ \langle \delta(q_{\alpha\beta} -
         \frac{1}{N}\sum_i w_i^\alpha w_i^\beta)\delta(R_\alpha -
         \frac{1}{N}\sum_i w_i^\alpha)\delta(R_\beta -
         \frac{1}{N}\sum_i w_i^\beta)\rangle }{\langle \delta(R_\alpha
         - \frac{1}{N}\sum_i w_i^\alpha)\delta(R_\beta -
         \frac{1}{N}\sum_i w_i^\beta)\rangle }
\end{eqnarray}
where the angled brackets denote averages over $w_i^\alpha$ and
$w_i^\beta$.  The weights at each site are distributed according to,
\begin{equation}
        p(w_i) = \left(\frac{1 + W_i}{2}\right)\delta(w_i - 1) +
        \left(\frac{1 - W_i}{2}\right)\delta(w_i + 1) \label{p(W_i)}
\end{equation}
Here, $W_i$ is the mean weight per site, defined in
equation~(\ref{W_i}).

We consider the Fourier transform of
$p(q_{\alpha\beta}|R_\alpha,R_\beta)$ since this appears in the
appropriate generating function,
\begin{equation}
        \hat{\rho}(-\i t|R_\alpha,R_\beta) = \frac{\hat{\rho}(-\i
        t,R_\alpha,R_\beta)}{\hat{\rho}(0,R_\alpha,R_\beta)}
\end{equation}
Writing the delta functions as integrals and noting that one of the
integrals is removed by the Fourier transform, one finds (ignoring
multiplicative constants),
\begin{equation}
        \hat{\rho}(-\i t,R_\alpha,R_\beta) = \left\langle
        \int_{-\i\infty}^{\i\infty} \!\! \d y_\alpha \d y_\beta
        \exp(F) \right\rangle_{\{w_i^\alpha,w_i^\beta\}}
\end{equation}
\[
        F = - y_\alpha R_\alpha - y_\beta R_\beta +
        \frac{1}{N}\sum_{i=1}^N(y_\alpha w_i^\alpha + y_\beta
        w_i^\beta + t w_i^\alpha w_i^\beta)
\]
Each site decouples and the average over sites can be taken by
integrating over the weight distribution defined in
equation~(\ref{p(W_i)}). The resulting integral can be computed for
large $N$ by the saddle point method since the exponent can be made
extensive by appropriate rescaling. Eventually one finds (ignoring
multiplicative constants),
\begin{equation}
        \hat{\rho}(-\i t,R_\alpha,R_\beta) = \exp(-y_\alpha R_\alpha -
        y_\beta R_\beta + G)
\end{equation}
\[
\fl G = \frac{1}{N} \sum_{i=1}^N
\log\left[(1+W_i)^2\e^{t+y_\alpha+y_\beta} +
2(1-W_i^2)\e^{-t}\cosh(y_\alpha-y_\beta) + (1 - W_i)^2
\e^{t-y_\alpha-y_\beta}\right]
\]
The saddle point equations fix $y_\alpha$ and $y_\beta$ as implicit
functions of $R_\alpha$, $R_\beta$ and $t$,
\begin{equation}
        R_\alpha = \frac{\partial G}{\partial y_\alpha} \qquad R_\beta
        = \frac{\partial G}{\partial y_\beta}
\end{equation}
Define $\hat{\rho}(-\i t)$, whose logarithm is the generating function
for $q_\infty$,
\begin{eqnarray}
        \hat{\rho}(-\i t) & = & \int \!\!\d R_\alpha\, \d R_\beta \,
        p_{\rm s}(R_\alpha) \, p_{\rm s}(R_\beta) \, \hat{\rho}(-\i
        t|R_\alpha,R_\beta) \nonumber \\ & = & \int \!\!\d
        R_\alpha\,\d R_\beta\, p_{\rm s}(R_\alpha) \, p_{\rm
        s}(R_\beta) \exp[G(t) - G(0)]
\end{eqnarray}
We express the overlap distributions by their Fourier transformed
cumulant expansions,
\begin{eqnarray}
        p_{\rm s}(R_\alpha) & = & -\i\int_{-\i\infty}^{\i\infty} \!\!
        \!\! \d a \,\exp\!\left(\sum\frac{a^n}{n!}K_n^{\rm s} -
        aR_\alpha\right) \\ p_{\rm s}(R_\beta) & = &
        -\i\int_{-\i\infty}^{\i\infty} \!\! \!\! \d b
        \,\exp\!\left(\sum\frac{b^n}{n!}K_n^{\rm s} - bR_\beta\right)
\end{eqnarray}
Now $\hat{\rho}(-\i t)$ is an integral over $a$, $b$, $R_\alpha$ and
$R_\beta$ which can again be computed by the saddle point method. One
finds that as $t \rightarrow 0$, the saddle point equations are
satisfied by,
\begin{eqnarray}
        y_\alpha \; & = & \; y_\beta \,\; = \,\; y \\ R_\alpha \: & =
        & \: R_\beta \; = \; K_1^{\rm s}
\end{eqnarray}
These are related through an implicit function for $y$ in terms of
mean overlap after selection,
\begin{equation}
        K_1^{\rm s} \; = \; \frac{1}{N}\sum_{i=1}^N \frac{W_i +
        \tanh(y)}{1 + W_i\tanh(y)}
\end{equation}
Then the natural increase contribution for the correlation after
selection is given by,
\begin{eqnarray}
        q_\infty & = & \lim_{t \rightarrow 0} \frac{\partial}{\partial
        t}\log\hat{\rho}(-\i t) \nonumber \\ & = &
        \frac{1}{N}\sum_{i=1}^N \left(\ \frac{W_i + \tanh(y)}{1 +
        W_i\tanh(y)} \right)^2
\end{eqnarray}

\end{document}